\documentclass[11pt]{amsart}

\usepackage{amssymb,amsmath,bm}
\usepackage{graphicx}
\usepackage[dvips]{color}
\usepackage[top=26mm, bottom=26mm, left=32mm, right=32mm]{geometry}

\newcommand{\vct}[1]{\bm{#1}}
\newcommand{\mtx}[1]{\mathsf{#1}}

\numberwithin{equation}{section}
\numberwithin{figure}{section}
\theoremstyle{definition}
\newtheorem{remark}{Remark}
\numberwithin{remark}{section}
\newtheorem{definition}{Definition}
\numberwithin{definition}{section}

\newcommand{\vtwo}[2]{\left[\begin{array}{c} #1 \\ #2 \end{array}\right]}

\newcommand{\mtwo}[4]{\left[\begin{array}{cc}    #1 & #2 \\ #3 & #4  \end{array}\right]}

\newcommand{\mbf}[1]{{\bm #1}}           
\newcommand{\infint}{\int_{-\infty}^{\infty} \!\!}      
\newcommand{\bi}{\begin{itemize}}
\newcommand{\ei}{\end{itemize}}
\newcommand{\ben}{\begin{enumerate}}
\newcommand{\een}{\end{enumerate}}
\newcommand{\be}{\begin{equation}}
\newcommand{\ee}{\end{equation}}
\newcommand{\bea}{\begin{eqnarray}}
\newcommand{\eea}{\end{eqnarray}}
\newcommand{\bc}{\begin{center}}
\newcommand{\bfi}{\begin{figure}}
\newcommand{\efi}{\end{figure}}

\newcommand{\ec}{\end{center}}
\newcommand{\tbox}[1]{{\mbox{\tiny #1}}}
\newcommand{\pO}{{\partial\Omega}}
\newcommand{\om}{\omega}
\newcommand{\al}{\alpha}

\newcommand{\ui}{u^\tbox{i}}                   
\newcommand{\uni}{u_n^\tbox{i}}                
\newcommand{\us}{u}                   
\newcommand{\RR}{\mathbb{R}^2}
\newcommand{\kx}{\kappa}                      
\newcommand{\kxi}{\kappa^\tbox{i}}                      
\newcommand{\ki}{k^\tbox{i}}                            
\newcommand{\xx}{\textbf{x}}             
\newcommand{\yy}{\textbf{y}}
\newcommand{\zz}{\textbf{z}}
\newcommand{\nn}{\textbf{n}}
\newcommand{\kk}{\textbf{k}}
\newcommand{\dd}{\textbf{d}}
\DeclareMathOperator{\sign}{sign}
\newcommand{\Sh}{\hat{\mathcal S}}
\newcommand{\Dh}{\hat{\mathcal D}}
\newcommand{\vt}[2]{\left[\begin{matrix}#1\\#2\end{matrix}\right]}
\newcommand{\mt}[4]{\left[\begin{matrix}#1&#2\\#3&#4\end{matrix}\right]}

\begin{document}
\title{A fast direct solver for quasi-periodic scattering problems}
\author{A. Gillman,  A. Barnett}
\address{Department of Mathematics, Dartmouth College}
\begin{abstract}
We consider the numerical solution of the scattering of time-harmonic
plane waves from an infinite periodic array of reflection or transmission
obstacles in a homogeneous background medium, in two dimensions.
Boundary integral formulations are ideal since they
reduce the problem to $N$ unknowns on the obstacle boundary.
However, for complex geometries and/or higher frequencies 
the resulting dense linear system becomes large,
ruling out dense direct methods,
and often ill-conditioned (despite being 2nd-kind),
rendering fast multipole-based iterative schemes also inefficient.
We present an
integral equation based solver with $O(N)$ complexity,
which handles such ill-conditioning,
using recent advances in ``fast'' direct linear algebra
to invert hierarchically the isolated obstacle matrix.
This is combined with a recent periodizing scheme
that is robust for all incident angles, including Wood's anomalies,
based upon the free space Green's function kernel.
The resulting solver is extremely
efficient when multiple incident angles are needed, as 
occurs in many applications.
Our numerical tests include
a complicated obstacle several wavelengths in size, with $N=10^5$
and solution error of $10^{-10}$,
where the solver is 66 times faster per incident angle
than a fast multipole based iterative solution,
and 600 times faster
when incident angles are chosen to share Bloch phases.
\end{abstract}

\maketitle

\section{Introduction}
\label{sec:intro}
Numerical modeling of the scattering of linear time-harmonic waves from
materials with periodic geometry plays a key
role in modern optics, acoustics, signal processing, and antenna
design.
Periodic diffraction problems where numerical modeling is crucial
include the design of
gratings for high-power lasers \cite{NIF},
thin-film solar cells \cite{atwater} and absorbers \cite{absorber},
process control in semiconductor lithography \cite{scatterometry},
linear water-wave scattering from pillars \cite{peter06},
and radar sensing of ocean waves.
In all such problems, an incident plane wave
generates a scattered wave which (in the far field)
takes the form of a finite sum of
plane waves at known Bragg angles, whose amplitudes are desired.

Several challenges arise in the efficient numerical solution of
grating scattering problems:
(i) the period of the gratings can be many wavelengths in size,
(ii) in many applications (such as photovoltaic \cite{atwater} or
solar absorber design \cite{absorber}) solutions are needed at many incident angles
and/or frequencies,
(iii) the scatterer may have physical resonances in the form of
guided modes (see Remark~\ref{r:guide}), which 
leads to ill-conditioned problems,
and
(iv) so-called Wood's anomalies may occur, that is,
scattering parameters (incident angle and frequency)
for which one of the Bragg angles lies precisely along the grating.
Note that challenges (iii) and (iv) are distinct:
(iii) is a physical resonance leading to an ill-posed problem
(see the reviews \cite{shipmanreview,lintonrev}),
whereas Wood's anomalies do not cause ill-posedness---and 
yet they do cause problems for many numerical schemes.

\begin{figure}[ht]
\centering
\setlength{\unitlength}{1mm}
\begin{picture}(120,100)
\put(-12,40){\includegraphics[width=70mm]{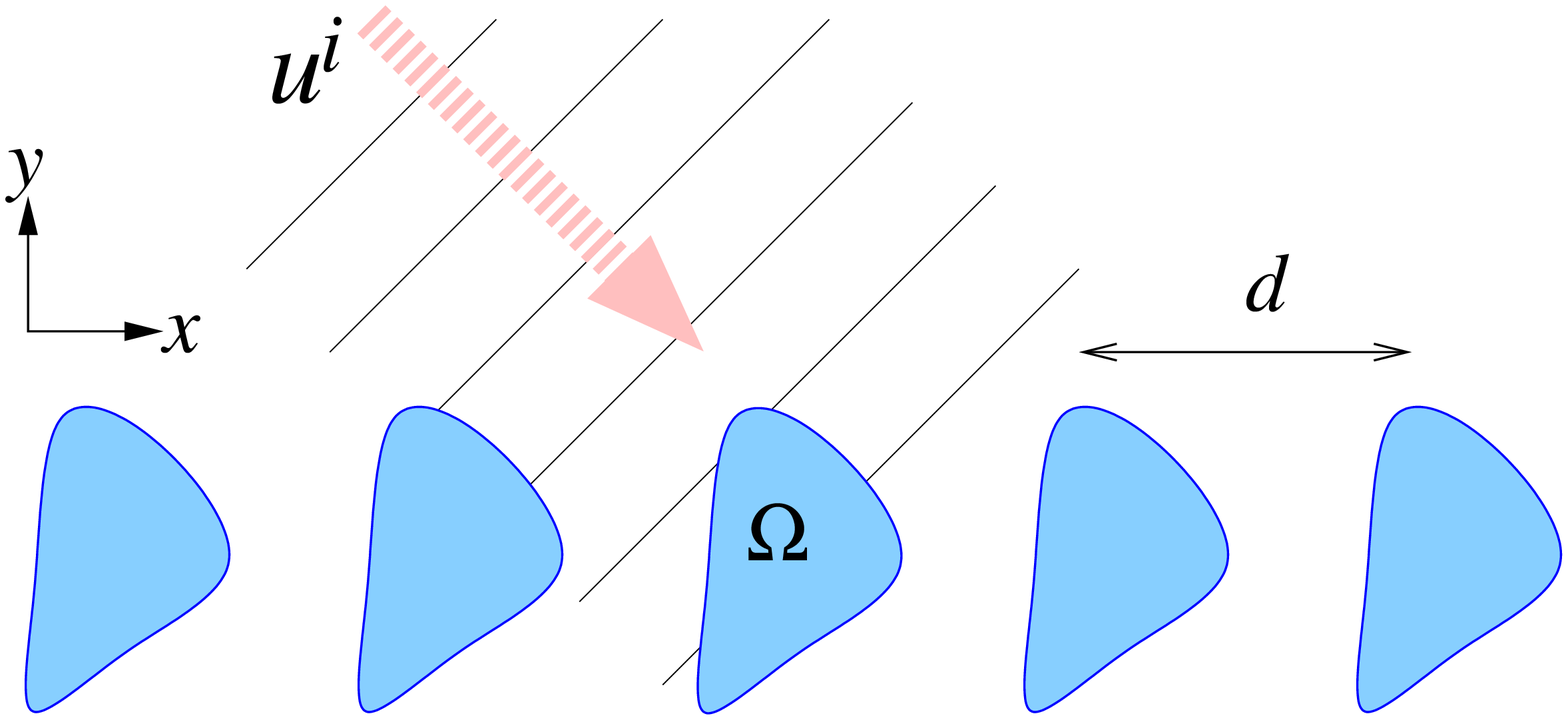}}
\put(70,40){\includegraphics[width=70mm]{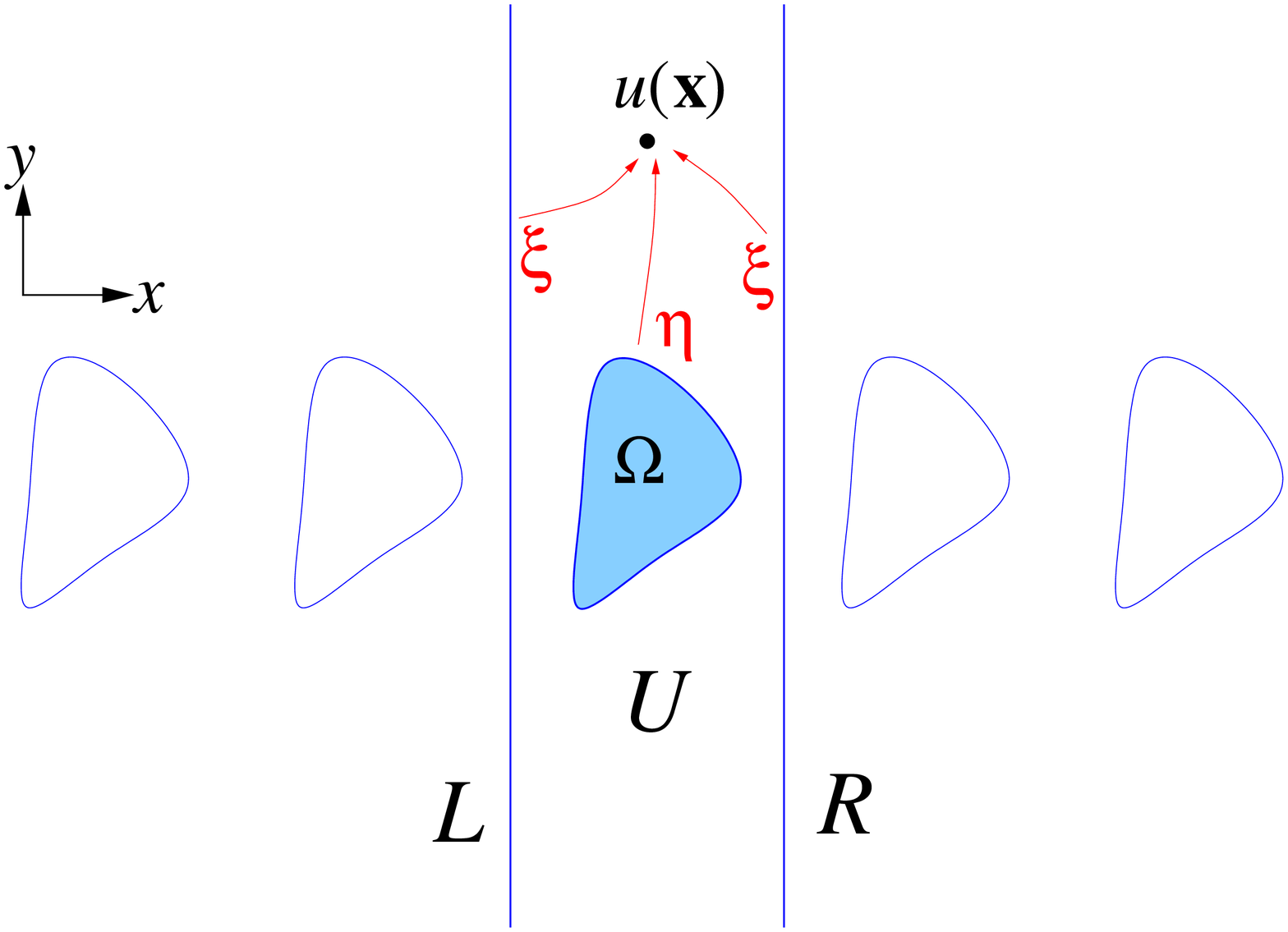}}
\put(10,05){\includegraphics[width=100mm]{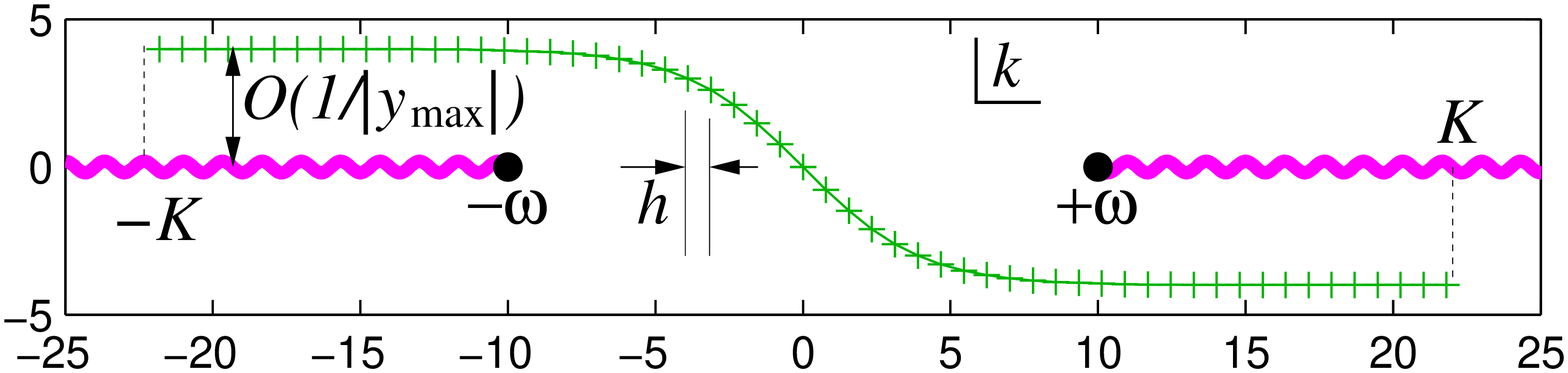}}
\put(15,37){(a)}
\put(100,37){(b)}
\put(00,20){(c)}
\end{picture}
\caption{\label{fig:geom}(a) Geometry of the periodic scattering problem.
(b) Geometry of the periodic scattering problem with artificial unit-cell
walls $L$ and $R$ introduced.  (c) Quadrature scheme on the Sommerfeld
contour in the complex Fourier plane (wiggly lines show branch cuts of the
integrand).}
\end{figure}

In most problems of interest, gratings consist of homogeneous media
delineated by sharp interfaces; hence the corresponding
partial differential equations (PDEs) have piecewise-constant coefficients.
This manuscript focuses on the
following two-dimensional Dirichlet boundary value problem,
which models acoustics with sound-soft obstacles,
or electromagnetic scattering
in TM (transverse magnetic) polarization from perfect electric conductors
(where $u$ represents the out-of-plane electric field \cite{qpsc}).
We seek the scattered wave $u$ which solves,
\begin{equation} \begin{split}
 (\Delta + \omega^2) u(\xx) &= 0 \ \ \qquad \qquad \qquad \xx := (x,y) \in \mathbb{R}^2\setminus \overline{\Omega_{\mathbb{Z}}}\\
 u (\xx)& = -u^{\rm i} (\xx)  \qquad \qquad \xx\in \partial \Omega_{\mathbb{Z}}\\
u & \quad\mbox{radiative as } \ \
y\to\pm\infty
~,
\end{split}
\label{eq:dir}
\end{equation}
where $\Omega\subset \mathbb{R}^2$ is a single bounded obstacle
which is repeated to form an infinite grating of obstacles
(denoted by $\Omega_{\mathbb{Z}}$)
of period $d$ along the $x$-axis (see Fig.~\ref{fig:geom}(a)).
A plane wave with frequency $\omega$, angle $\theta \in(-\pi,0)$, and wavevector
$\kk = (\kxi,\ki)  = (\om \cos \theta, \om \sin\theta)$
is incident; hence $u^{\rm i} (\xx)= e^{i\kk\cdot \xx}$
outside the obstacles and vanishes inside.
It is easy to see from \eqref{eq:dir} that
the total field $u^t = u^{\rm i}+u$ vanishes on $\pO$:
this is the physical boundary condition.
For the exact radiation condition on the scattered field
$u$ see \eqref{rbu}-\eqref{rbd}.
The incident field, and hence the scattered field, satisfy
a \textit{quasi-periodicity} condition,
\begin{equation}
u(x+d,y) = \alpha u(x,y) \qquad \forall (x,y) \in \mathbb{R}^2
~,
\label{qp}
\end{equation}
where $\alpha := e^{i\kxi d} = e^{i \omega d\cos \theta}$ is known as the Bloch phase.
We will also consider the transmission problem
corresponding to dielectric obstacles (see section~\ref{sec:trans}).

Integral equation formulations are a natural choice 
for problems with piecewise-constant coefficients such as \eqref{eq:dir}:
they exploit this fact by reducing the problem to
an integral equation whose discretization involves
$N$ unknowns living on the boundary alone.
This has many advantages over finite elements or finite difference schemes:
it reduces the dimensionality by one,
allowing for more complicated boundaries to be solved to high accuracy
with a small $N$, and is amenable to easily implemented high-order quadratures.

In the grating application, the infinite boundary $\partial \Omega_{\mathbb{Z}}$
must be reduced to the boundary of a {\em single} obstacle $\pO$;
there are two main approaches to this task of ``periodizing'' the integral
equations.

The first approach replaces the Green's function (fundamental solution)
that appears in
the kernel of in the integral operators by its quasi-periodized version
\cite{lintonrev} (which satisfies \eqref{qp}).
This is used in two dimensions by
Bruno--Haslam \cite{brunohaslam09} and in three by Nicholas \cite{nicholas}
and Arens \cite{arenshabil}.
However, this fails as parameters approach a Wood's anomaly,
 because the Green's function does not exist there.
One cure (in the case of a connected interface) is
to use the quasi-periodic impedance half-space
Green's function \cite{CWnystrom,arens06}.
However, all such
periodized kernel methods do not scale well to large $N$ because
dense matrices must be filled at a cost of order a millisecond
per element \cite{horoshenkov}.

The second approach,
which is robust at all parameters including at or near Wood's anomalies,
is to return to the free-space Green's function
and instead periodize using a small number of additional unknowns on artificial
unit-cell walls.
The condition \eqref{qp} is then imposed directly in the linear system.
This is used by Wu--Lu \cite{wulu09},
and also by one of the authors
in \cite{qpsc}.
Further advantages of the second approach include: the {\em low-rank} nature of the
periodizing is explicit (no dense matrices of
periodized Green's evaluations are needed),
and that
free-space boundary integral equation quadratures and codes may be used without modification.

In this paper, we present a fast direct solver for the formulation in
\cite{qpsc}, that can handle, in reasonable computation times,
complicated boundaries that both demand large $N$
and cause ill-conditioning.
Section \ref{sec:int} reviews the derivation of the
periodized integral formulation in the
Dirichlet setting \eqref{eq:dir}.

Like many boundary integral equations for two-dimensional domains,
this gives a linear system that could be solved
via an iterative solver (e.g.\ GMRES) coupled with a fast matrix-vector multiplication scheme such as
the fast multipole method (FMM) \cite{fmm2}.
Unfortunately, when the system is ill-conditioned (which often occurs for complicated geometry) it can take hundreds of iterations for 
such an iterative solver to converge, or worse, it may never converge to an acceptable accuracy.
Additionally, iterative methods are not able to efficiently solve problems 
with multiple right hand sides that occur when the response is needed at multiple incident angles.
This has motivated in recent years the development of a collection of
{\em fast direct solvers}.
These solvers utilize internal structure (essentially low-rank off-diagonal blocks) to 
construct rapidly an approximate inverse of the dense matrices resulting from the the discretization of integral equations.
For many problems, the computational cost is $O(N\log ^k N)$
with typically $k =0,1,$ or $2$ 
\cite{hackbusch,2010_borm,mdirect,BCR,1996_mich_elongated,1994_starr_rokhlin},
which can be orders of magnitude smaller than traditional
$O(N^3)$ dense direct methods.

In addition to being fast, the solvers are robust and construct the approximate inverse in such a way that it 
can be applied in with $O(N)$ (with small constant) computational cost. 
In this work, we choose a Hierarchically Block Separable (HBS) solver \cite{m2011_1D_survey}.  Like the method in \cite{ho}, 
the solver, described in section \ref{sec:HSS}, exploits potential theory to create low-rank factorizations.  As a result, the computational cost scales 
linearly with the number of discretization points for low-frequency problems on many
domains.

Upon discretization, the integral formulation of the grating scattering problem
in section \ref{sec:int}
gives a two-by-two block linear system. Section \ref{sec:direct}
presents an efficient technique 
for solving this
system with multiple right-hand sides, assuming an inverse of the large
$N\times N$ block is available.
The full fast direct scheme for the problem is then achieved by
using the HBS solver of section \ref{sec:HSS} to compute and apply this
large matrix inverse whenever it is needed in the technique of
section \ref{sec:direct}.

We describe how our technique may be adapted to the transmission scattering problem in 
section \ref{sec:trans}.
Section \ref{sec:numerics}
illustrates the performance of the fast direct solver in some
test cases,
and compares the computation time to a standard fast iterative scheme.
Finally, we draw some conclusions in section \ref{sec:conc}.

\section{Integral formulation for the quasi-periodic Dirichlet problem}
\label{sec:int}
In this section, we describe an integral equation formulation for the
grating scattering problem \eqref{eq:dir} that is based upon \cite{qpsc}.
First we rephrase this boundary-value
problem as an equivalent one on the unit cell containing a single obstacle.

\subsection{Restriction to a problem on a single unit cell}
\label{s:restr}

We assume that the copies of $\Omega$ do not intersect and
that we can create a unit cell ``strip'' $U:=\{(x,y): |x|<d/2, y\in\mathbb{R}\}$ containing the closure of 
$\Omega$.  Let the infinite bounding left and right walls be denoted by $L$ and $R$ respectively
(see Figure~\ref{fig:geom}(b)).
As in \cite{bonnetBDS}, the boundary value problem can be rephrased as an equivalent problem on the domain $U$, that is
\bea
(\Delta + \om^2) u & = & 0 \qquad \mbox{ in } U\backslash \overline{\Omega}
\label{pde}
\\
u & = & -\ui \quad \mbox{ on } \pO
\label{bc}
~.
\eea
Quasi-periodicity of the field is imposed via wall matching conditions,
\bea
u|_L - \al^{-1}u|_R &=& 0
\label{f}
\\
u_n|_L - \al^{-1}u_n|_R &=& 0
\label{fn}
\eea
where the normal derivatives on $L$, $R$ are taken to be in the positive
$x$ direction.
The total field $u$ satisfies the outgoing radiation condition
\cite{bonnetBDS}, which is to say that it
is given by uniformly-convergent Rayleigh--Bloch expansions,
\bea
\us(x,y) &=& \sum_{n\in\mathbb{Z}} c_n
   e^{i \kx_n x} e^{i k_n (y-y_0)}
   \qquad y>y_0, \; |x|\le d/2  
\label{rbu}
\\
\us(x,y) &=& \sum_{n\in\mathbb{Z}} d_n
   e^{i \kx_n x} e^{i k_n (-y-y_0)}
   \qquad y<-y_0, \; |x|\le d/2         
\label{rbd}
\eea
where $y_0>\sup_{(x,y)\in\Omega}|y|$,
so that $\overline\Omega$ lies within the vertical bounds
$|y|<y_0$,
the $x$-component of the $n$th mode wavevector is
$\kappa_n := \kxi + 2\pi n/d$, and the $y$-component is
$k_n := +\sqrt{\om^2-\kappa_n^2}$.
The sign of the square-root is taken as non-negative real or positive imaginary.
The coefficients $c_n$ and $d_n$ for orders $n$ that are propagating ($|\kx_n|\le \om$) 
are the desired amplitudes of the Bragg orders mentioned in the Introduction.

The above radiation condition also completes the precise description
of the scattering problem \eqref{eq:dir}.
We now make a remark about well-posedness of \eqref{eq:dir},
which of course equally well applies to the single unit cell
version presented above.

\begin{remark} 
With the radiation condition \eqref{rbu}--\eqref{rbd},
at least one solution exists to \eqref{eq:dir} at all
scattering parameters (angles $\theta$ and frequencies $\om$) \cite[Thm.~3.2]{bonnetBDS}.
Furthermore, at each angle, the solution is unique except possibly at a discrete set of
frequencies which correspond to physical guided modes.
The only such modes accessible in the scattering setting
are embedded in the continuous spectrum \cite{shipmanreview}.
\cite{bonnetBDS,shipmanreview} give conditions for nonexistence
of such modes are given.
These results will also hold for the
transmission version where \eqref{pde}--\eqref{bc} are replaced
by \eqref{pdein}--\eqref{bcmn} below.
\label{r:guide}
\end{remark}


\subsection{An indirect boundary integral equation formulation}
\label{s:bie}

Recall the single- and double-layer potentials \cite{CK83}
on an obstacle boundary $\pO$, defined as
\begin{equation}
({\mathcal S}_\pO\sigma)(\xx) = \int_\pO G(\xx,\yy) \sigma(\yy) ds_\yy,
\label{S}
\end{equation}
and
\begin{equation}
({\mathcal D}_\pO\tau)(\xx) = \int_\pO
\frac{\partial G}{\partial \nn(\yy)}(\xx,\yy) \tau(\yy) ds_\yy,
\label{D}
\end{equation}
respectively.
The kernel $G(\xx,\yy):={\mathcal G}(\xx-\yy)$
is the free-space Green's function for the Helmholtz equation at frequency $\omega$,
\be
{\mathcal G}(\xx) = \frac{i}{4}H_0^{(1)}(\om|\xx|),
\qquad\xx\in\RR\setminus\{\mbf{0}\},
\label{G}
\ee
where $H_0^{(1)}$ is the outgoing Hankel function of order zero.

For the purposes of periodizing, auxiliary layer potentials on the
$L$ and $R$ walls are needed. To handle their infinite extent, we 
switch from coordinate $y$ to the Fourier variable $k$, via
\be
\hat{g}(k) = \frac{1}{2\pi} \infint e^{-iky} g(y) dy\, ,
\qquad
g(y) = \infint e^{iky} \hat{g}(k) dk \, ,
\label{ft}
\ee
and make use of the spectral representation of the free-space Green's function
\cite[Ch. 7.2]{m+f},
\be
{\mathcal G}(\xx) \;=\; \frac{i}{4\pi}\infint
e^{iky}
\;\frac{e^{i\sqrt{\om^2-k^2}\, |x|}}{\sqrt{\om^2-k^2}} \; dk
\, , \qquad \xx=(x,y)
\label{specrep}
\ee

Square-roots are again taken non-negative real or positive imaginary,
achieved by taking the branch
cuts of the function $\sqrt{\om^2-k^2}$ in the $k$ plane as
$(-\infty, -\om) \cup (\om,+\infty)$ along the real axis
and using a so-called Sommerfeld contour \cite{chewbook} for the integration passing
from the 2nd to 4th quadrants (Fig \ref{fig:geom}(c)).
Inserting \eqref{specrep} into the usual expressions for single- and double-layer potentials
living on a vertical wall $W=\{(x_0,y):y\in\mathbb{R}\}$
($W$ will be $L$ or $R$, where $x_0$ takes the values $-d/2$ or $d/2$
respectively), we get
``Fourier layer potentials''
\bea
(\Sh_W\hat{\mu})(\xx) & =& \frac{i}{2} \infint e^{iky}
e^{i\sqrt{\om^2-k^2}\, |x-x_0|}
\frac{1}{\sqrt{\om^2-k^2}} \hat{\mu}(k) \,dk
~,
\label{Sh}
\\
(\Dh_W\hat{\nu})(\xx) & =& \frac{\sign(x-x_0)}{2} \!\!
\infint \!e^{iky}
e^{i\sqrt{\om^2-k^2}\, |x-x_0|}
\hat{\nu}(k) \,dk
~.
\label{Dh}
\eea
Here $\hat{\mu}$ and $\hat{\nu}$ are interpreted as Fourier-transformed layer densities,
or the coefficients of a plane-wave representation.

We use a standard combined-field representation
(to avoid spurious interior obstacle resonances \cite{CK83})
with density $\eta$ on the obstacle boundary,
and auxiliary densities $\hat\xi=[\hat\mu;\hat\nu]$ which
will represent fields due to the remaining lattice of obstacles in the
grating, thus
\be
u = ({\mathcal D}_\pO -i\om{\mathcal S}_\pO)\eta \,+\,
(\Sh_L + \al \Sh_R)\hat\mu + (\Dh_L + \al \Dh_R)\hat\nu
\qquad \mbox{ in } U\backslash\overline{\Omega}
\label{urep}
\ee
See Fig.~\ref{fig:geom}(b).
Imposing the Dirichlet boundary condition in \eqref{eq:dir} whilst
the imposing quasi-periodicity at the walls \eqref{f}--\eqref{fn}
results in the first and second rows, respectively, of the following 2-by-2 linear operator
system 
\be
\mt{A}{\hat B}{\hat C}{\hat Q}\vt{\eta}{\hat\xi} = \vt{ -\ui|_\pO}{0}
~,
\label{E}
\ee
where the four operators $A$, $\hat B$, $\hat C$ and $\hat Q$ are defined in the remainder 
of this section.
\footnote{Note that operators with the symbol $\land$ involve Fourier
variables, a notation consistent with \protect\cite{qpsc}.}

Using jump relations \cite{CK83}, we have $A = I/2 + D - i\om S$ where $S, D: C(\pO)\to C(\pO)$ are the boundary
operators with the kernels of \eqref{S} and \eqref{D}, respectively.

The operator $\hat B$ gives the effect of the auxiliary densities on the obstacle
boundary value. By restricting \eqref{Sh}--\eqref{Dh} and \eqref{urep} to $\partial \Omega$,
\be
\hat B = [\Sh_{\pO,L} + \al \Sh_{\pO,R}, \;\; \Dh_{\pO,L} + \al \Dh_{\pO,R}]
~,
\label{B}
\ee
where $\Sh_{\pO,W}$ and $\Dh_{\pO,W}$ denote the operators resulting from restricting
\eqref{Sh} and \eqref{Dh} to evaluation on $\pO$.

Following \cite{qpsc}, we impose the second row of \eqref{E} in Fourier space to
enable an efficient discretization with spectral accuracy.  
Via \eqref{S}, \eqref{D} and \eqref{specrep}, 
the operators mapping single- and double-layer densities on $\pO$
to their Fourier values on a wall $W$ are defined by
$$
(S^\land_{W,\pO}\sigma) (k) = \frac{i}{4\pi}\int_\pO e^{-iky}
\frac{e^{i\sqrt{\om^2-k^2}\, |x-x_0|}}{\sqrt{\om^2-k^2}} \,\sigma(\yy) ds_\yy,
\qquad \yy=(x,y)\in\pO
\nonumber 
$$
and 
$$
(D^\land_{W,\pO}\tau) (k) = \frac{1}{4\pi}\int_\pO e^{-iky}
e^{i\sqrt{\om^2-k^2}\, |x-x_0|}
\Bigl(-\sign(x-x_0),\frac{k}{\sqrt{\om^2-k^2}}\Bigr)\cdot \nn(\yy)
\, \tau(\yy) ds_\yy
~.
\nonumber
$$
Similarly, the following two operators which
instead map to Fourier normal derivatives on the wall $W$ are defined by
\bea
(D^{\ast,\land}_{W,\pO}\sigma) (k) &=& \frac{1}{4\pi}\int_\pO e^{-iky}
e^{i\sqrt{\om^2-k^2}\, |x-x_0|} \sign(x-x_0)
\, \sigma(\yy) ds_\yy
\nonumber \\
(T^\land_{W,\pO}\tau) (k) &=& \frac{i}{4\pi}\int_\pO e^{-iky}
e^{i\sqrt{\om^2-k^2}\, |x-x_0|}
\bigl(\sqrt{\om^2-k^2},-k\sign(x-x_0)\bigr)\cdot \nn(\yy)
\, \tau(\yy) ds_\yy
~.
\nonumber
\eea

These four formulae allow us to express the operator $\hat C$ in \eqref{E},
which gives the effect of the density $\eta$ on the Fourier
transforms of the wall quasi-periodicity
conditions \eqref{f}-\eqref{fn}, as
\be
\hat C = \vt{D^\land_{L,\pO} - i\om S^\land_{L,\pO} - \al^{-1}
(D^\land_{R,\pO} - i\om S^\land_{R,\pO})}
{T^\land_{L,\pO}- i\om D^{\ast,\land}_{L,\pO} - \al^{-1}
(T^\land_{R,\pO} - i\om D^{\ast,\land}_{R,\pO})}
~.
\label{C}
\ee

Finally, $\hat Q$ maps Fourier wall densities to Fourier
wall  quasi-periodicity conditions. Thus, by translational
invariance, each of its four blocks must be a pure multiplication operator in $k$.
The Fourier coefficients of \eqref{Sh}--\eqref{Dh} in \eqref{urep}
and \eqref{f}--\eqref{fn} give,
\be
\hat Q = \mt{I}{0}{0}{I} + \frac{e^{i\sqrt{\om^2-k^2}\,d}}{2}
\mt{i(\al-\al^{-1})/\sqrt{\om^2-k^2}}
{-\al -\al^{-1}}{\al+\al^{-1}}{i(\al-\al^{-1})\sqrt{\om^2-k^2}}
~.
\label{Q}
\ee

\begin{remark}
\label{r:bragg}
Once the operator system \eqref{E} is solved for $\eta$ and $\hat\xi$, the
desired Bragg amplitudes $c_n$ and $d_n$
appearing in \eqref{rbu}--\eqref{rbd}
are extracted by evaluating the scattered field $u$ in \eqref{urep},
and its $y$-derivative,
at typically 20
equi-spaced samples along the lines $\{(x,y): |x|\le d/2, y=\pm y_0\}$,
then applying the discrete Fourier transform (e.g. via the FFT).
See \cite{qpsc} for more details.
\end{remark}


\subsection{Discretizing the integral equations}
\label{s:quad}
This section describes a discretization of the 
linear operator system (\ref{E}) with high-order accuracy
to obtain a finite-sized linear system.
We discretize all operator blocks of $E$ via the Nystr\"om method \cite{NA}.
Since the $A$ operator has a kernel with a logarithmic singularity
at the diagonal, it requires special quadrature corrections.
For this, we choose the 6th-order Kapur--Rokhlin rule \cite{Kapur};
however, the fast direct solver is compatible
with other recently-developed local-correction quadrature schemes
such as that of Alpert \cite{alpert}, that of Helsing \cite{helsing},
generalized Gaussian \cite{2011_hao_nystrom}, or QBX \cite{QBX}.

The boundary $\pO$ is parameterized by the smooth $2\pi$-periodic
function $\zz:[0,2\pi)\to\pO$, 
then discretized using the $N$-point global periodic trapezoid rule
with nodes $\yy_j = \zz(2\pi j/N)$, $j=1,\ldots,N$.
Then the $N$-by-$N$ matrix $\mtx{A}$ represents the operator $A$.
The Kapur--Rokhlin scheme modifies the weights (but not the nodes)
near the diagonal, giving the matrix entries
$$\mtx{A}_{nm} = (2\pi/N) R_{|n-m|} [\partial G(\yy_n,\yy_m)/\partial \nn(\yy_m) - i\om G(\yy_n,\yy_m)]
|\zz'(2\pi m/N)|.$$
The values $R_j$ are given
in terms of $\gamma_j$
from the left-center block of \cite[Table 6]{Kapur},
as follows:
$R_0 = 0$, while $R_j = R_{N-j} = 1+\gamma_j+\gamma_{-j}$ for $1\le j\le6$,
and $R_j = 1$ otherwise.

There is some freedom in choosing a contour for the
Sommerfeld $k$-integrals in \eqref{Sh}--\eqref{Dh}.  
We choose a hyperbolic tangent curve of height $O(1)$ and use
the trapezoid rule in the real part of $k$ with spacing $h$,
truncated to maximum real part $K$
(see Fig.~\ref{fig:geom}(b) and \cite{qpsc}).
$K$ is chosen such that the integrand is
exponentially small beyond this real part.
There are $M = 2K/h$ nodes $k_j\in\mathbb{C}$ for $j=1,\dots,M$.
Since the scheme is exponentially convergent, $M$ is typically only 100 for full machine
precision when $\om$ is around 10.
Note that since there are two types of auxiliary densities, there are $2M$
unknowns in the periodizing scheme.
Using the above quadrature nodes and weights,
$\mtx{B}$ and $\mtx{C}$ are simply Nystr\"om discretizations of \eqref{B} and
\eqref{C}, while
$\mtx{Q}$ has four diagonal sub-blocks with entries given by \eqref{Q}
evaluated at the Sommerfeld nodes.
The entire square block system in $N+2M$ unknowns is written
(we drop the $\land$ symbols from now on for simplicity),
\begin{equation}\mtwo{\mtx{A}}{\mtx{B}}{\mtx{C}}{\mtx{Q}} \vtwo{\vct{\eta}}{\vct{\xi}} = \vtwo{\vct{b}}{0}\label{eq:block}
~,
\end{equation}
where $\vct{b}$ is the vector of values of $-\ui$ at the nodes on $\pO$.

\begin{remark}
\label{r:braggnyst}
Once $\vct{\eta}$ and $\vct{\xi}$ have been solved for,
the desired Bragg amplitudes may be computed as in Remark~\ref{r:bragg},
evaluating \eqref{urep}
using the underlying Nystr\"om quadratures on $\pO$ and
in the Fourier $k$ variable.
\end{remark}


\subsection{Improving the convergence rate, and robustness near Wood's anomalies}
\label{sec:improve}
In this section, we briefly sketch two ways to improve the convergence rate
and robustness of the
integral formulation (see \cite{qpsc} for full details).

Firstly, we include nearest neighbor images of the obstacle in the
representation, replacing \eqref{urep} with 
\be
u = \sum_{j=-P}^P \al^j({\mathcal D}_{\pO+j\dd}
-i\om{\mathcal S}_{\pO+j\dd})\eta \,+\,
(\Sh_L + \al \Sh_R)\hat\mu + (\Dh_L + \al \Dh_R)\hat\nu
\qquad \mbox{ in } U\backslash\overline{\Omega}
\label{urepP}
\ee
where $\dd=(d,0)$ is the lattice vector and $P$ is the number of nearest neighbors on either side of $\Omega$ to be included.
Since the auxiliary periodizing wall densities have to represent
fields whose nearest singularities are now further away,
this improves the convergence rate with respect to $M$.
We have found that $P=1$ or $P=2$ is optimal.
As a result, $A$ will now contain not only a self-interaction of $\pO$,
but interactions from the boundaries of $2P$ obstacles neighboring $\Omega$.

Recall that $\mtx{A}$ is already defined as the matrix from Nystr\"om discretization of the self-interaction operator
$A = I/2+D_{\pO,\pO} - i\om S_{\pO,\pO}$, where $D_{V,W}$ and $S_{V,W}$ represent
the integral operators with single- and double-layer kernels acting from source curve $W$ to target curve $V$.
For $j\in\mathbb{Z}\backslash\{0\}$, let $\mtx{A}_j$ 
denote the matrix resulting from Nystr\"om discretization of the operator $A_j = D_{\pO,\pO+j\dd} - i\om S_{\pO,\pO+j\dd}$.
This expresses the effect on $\pO$ of a neighbor $j$ obstacles from $\pO$.
Let $$\tilde{\mtx{A}} = \displaystyle\mtx{A}+\sum_{\substack{j=-P \\j\neq 0}}^P\alpha^j\mtx{A}_j.$$
Then, with this new representation, the block linear system (\ref{eq:block}) is 
replaced by 
\begin{equation}\mtwo{\tilde{\mtx{A}}}{\mtx{B}}{\tilde{\mtx{C}}}{\mtx{Q}} \vtwo{\vct{\eta}}{\vct{\xi}} = \vtwo{\vct{b}}{0}\label{eq:block2}
~,
\end{equation}
where $\tilde{\mtx{C}}$ is the Nystr\"om discretization of
\be
\tilde{ C} = \vt{\al^P(D^\land_{L,\pO+P\dd} - i\om S^\land_{L,\pO+P\dd}) - \al^{-(P+1)}
(D^\land_{R,\pO-(P+1)\dd} - i\om S^\land_{R,\pO-(P+1)\dd})}
{\al^P(T^\land_{L,\pO+P\dd}- i\om D^{\ast,\land}_{L,\pO+P\dd}) - \al^{-(P+1)}
(T^\land_{R,\pO-(P+1)\dd} - i\om D^{\ast,\land}_{R,\pO-(P+1)\dd})}
~.
\nonumber 
\ee
which results from the new representation
via some cancellations as in \cite{qpsc}.

The scheme as presented so far would fail as one approaches a Wood's anomaly,
since the auxiliary densities $\hat\mu(k)$ and $\hat\nu(k)$ contain
poles at $k=\pm k_n$, and as any $k_n$ approaches the origin
this makes the Sommerfeld quadrature highly inaccurate \cite[Sec.~5]{qpsc}.
Thus, the
second way to improve the scheme is to, in this situation,
displace the Sommerfeld contour
by an $O(1)$ real value such that it is far from all poles $\pm k_n$.
This causes an incorrect radiation condition for the Rayleigh--Bloch
mode whose pole the contour crossed. However, this is fixed simply by
including an unknown coefficient for this mode
in the representation 
\eqref{urepP}, and imposing
the one extra linear condition that the radiation condition \eqref{rbd}
be correct.
The elements of the matrix row which imposes the condition
that there be no {\em incoming} radiation in the relevant mode
are computed as in Remark~\ref{r:braggnyst}.

The net effect is that,
by expanding the system \eqref{eq:block}
or \eqref{eq:block2} by one extra row and column,
parameters at or near Wood's anomalies are also handled to machine precision.
This robustness
distinguishes the scheme from many other integral-equation-based solvers.


\section{A direct solution technique}
\label{sec:direct}
This section presents techniques for solving the discretized linear system \eqref{eq:block2} in the case of
multiple incident angles, using only a single inversion of the matrix
$\mtx{A}$.
Coupling these techniques with the method described 
in section \ref{sec:HSS} and \cite{gillman} will result in a fast direct solver for the Dirichlet
scattering problem \eqref{eq:dir}.

For simplicity, we start with the block solution technique assuming no
contributions from neighboring obstacles, i.e.\ $P=0$.  Then we describe how to handle $P>0$ by exploiting the internal structure of the matrices
characterizing the interaction between $\partial \Omega$ and its $2P$ nearest neighbors.

\subsection{The block solve (case $P=0$)}
\label{sec:block}
When there are no neighbor contributions,
the $2\times 2$ linear system \eqref{eq:block} has
solutions $\vct{\eta}$ and $\vct{\xi}$ given by
\begin{equation}\begin{aligned}
 \vct{\vct{\xi}} & = \left(\mtx{Q}-\mtx{C}\mtx{A}^{-1}\mtx{B}\right)^{-1}\mtx{A}^{-1} \vct{b}\\
\vct{\vct{\eta}} & = \mtx{A}^{-1} \vct{b}- \mtx{A}^{-1} \mtx{B}\vct{\xi}.
\end{aligned}\label{eq:blocksolve}
\end{equation}
Their computation involves inverting two matrices $\mtx{A}$ and $\left(\mtx{Q}-\mtx{C}\mtx{A}^{-1}\mtx{B}\right)$.
Of course this assumes that $\mtx{A}$ is invertible;
however, this is known to hold for sufficiently
large $N$ since it is the Nystr\"om discretization of the
injective operator $I/2+D-i\om S$ arising in the
scattering from an isolated obstacle \cite[p.~48]{coltonkress}.
Since the size of $\mtx{Q}$ is much less than $N$ (typically $M \sim 100$), the cost of the solve is dominated by the inversion 
of the $N\times N$ matrix $\mtx{A}$. Fortunately, $\mtx{A}$ has internal structure which makes it amenable to an $O(N)$ inversion 
technique described in section \ref{sec:HSS}.
Notice that while $\mtx{B}$, $\mtx{C}$ and $\mtx{Q}$ do depend upon the incident angle via their dependence on $\alpha$,
the matrix $\mtx{A}$ does not.  Thus $\mtx{A}$ may be inverted once and for all at each frequency $\omega$.

\begin{remark}
At a fixed frequency $\om$,
there may be multiple incident angles $\theta$
that share a common Bloch phase $\alpha$ via the relation
$\alpha = e^{i \omega d\cos \theta}$.
Let $q$ be the number of incident angles that share an $\alpha$.
It is then easy to check that, within $\pm 1$, we have
$q \approx \om d/\pi$,
so that $q$ is proportional to the grating period in wavelengths.
Notice that the number of matrix-vector
multiplies with $\mtx{A}^{-1}$ required to evaluate
\eqref{eq:blocksolve} is then $2M + q$, and that $\mtx{A}^{-1}$ need only 
be accessed twice. We will later exploit this fact for efficiency.
\label{r:multiangle}
\end{remark}

\subsection{The block solve with neighboring contributions}
\label{sec:neigh}
Recall from section \ref{sec:improve} that, when contributions from neighboring obstacles are included in the integral formulation,
the matrix $\mtx{A}$ in (\ref{eq:blocksolve}) is 
replaced by $$\tilde{\mtx{A}}= \mtx{A}+\sum_{\substack{j = -P,\\j\neq 0}}^{P}\alpha^j \mtx{A}_j.$$
In practice, we find $P=1$ or 2 is best as $P>2$ gives little additional
accuracy.

Because $\pO$ is separated from its neighbors, the
matrices $\mtx{A}_j$ are low rank (i.e. $\mtx{A}_j$ has rank $l$ where $l\ll N$). Thus they admit 
a factorization $\mtx{A}_j = \mtx{L}_j\mtx{R}_j$ where $\mtx{L}_j$ and $\mtx{R}_j$ are of size $N\times l$.  This means that 
the matrix that needs to be inverted is $\mtx{A}+\mtx{L}\mtx{R}$ where 
$$\mtx{L} = \left[\alpha^{-P}\mtx{L}_{-P} | \cdots | \alpha^{-1}\mtx{L}_{-1} |  \alpha\mtx{L}_{1} | \cdots | \alpha^{P}\mtx{L}_{P}\right]$$
and
$$\mtx{R} = \left[\mtx{R}_{-P}^T |\dots |\mtx{R}_{-1}^T |\mtx{R}_1^T | \dots | \mtx{R}_{P}^T\right] ^T ~.$$

The inverse of $\tilde{\mtx{A}} = \mtx{A}+\mtx{L}\mtx{R}$ can be computed using only the inverse of $\mtx{A}$ via the Woodbury formula \cite{golub}
\begin{equation}\tilde{\mtx{A}}^{-1} = (\mtx{A} +\mtx{L}\mtx{R} )^{-1} = \mtx{A}^{-1} + \mtx{A}^{-1}\mtx{L}\left(\mtx{I}+\mtx{R}\mtx{A}^{-1}\mtx{L}\right)^{-1}\mtx{R}\mtx{A}^{-1}.\label{eq:inv}\end{equation}
Note that the square matrix $\left(\mtx{I}+\mtx{R}\mtx{A}^{-1}\mtx{L}\right)$
is only of size $2Pl$ thus can easily be inverted with dense linear
algebra.
Moreover, in practice we do not actually compute $\tilde{\mtx{A}}^{-1}$.
Instead we apply $\tilde{\mtx{A}}^{-1}$ via \eqref{eq:inv} which requires
two applications of $\mtx{A}^{-1}$.  For example, to multiply $\tilde{\mtx{A}}^{-1}$ by a vector $\vct{x}$, we need 
to evaluate $\mtx{A}^{-1} \mtx{L}$ and $\mtx{A}^{-1}\vct{x}$.  
For large $N$, this is done 
via the techniques described in section~\ref{sec:HSS}.
%

Instead of using a QR factorization to find $\mtx{L}_j$ and $\mtx{R}_j$, we choose to use an interpolatory decomposition \cite{gu1996,lowrank}
defined as follows.
\begin{definition}
 The \textit{interpolatory decomposition} of a $m\times n$ matrix $\mtx{M}$ that has rank $l$ is
the factorization 
$$ \mtx{M} = \mtx{P}\mtx{M}(J(1:l),:)$$
where $J$ is a vector of integers $j_i$ such $1\leq j_i\leq m$, and $\mtx{P}$ is a $m\times l$ matrix that contains a $l \times l$ identity matrix.
Namely, $\mtx{P}(J(1:l),:) = \mtx{I}_l$.  
\end{definition}

Since the $\mtx{L}_j$ and $\mtx{R}_j$ matrices do not involve $\alpha$, they need only be computed once at each frequency, independent of 
the number of incident angles.
However, the cost of computing these factorizations
using general linear algebraic techniques such as QR 
is $O(N^2 l)$.  This would negate the substantial savings obtained by using the $O(N)$ inversion technique for $\mtx{A}$ to be described 
in section \ref{sec:HSS}.

To restore the complexity, we use ideas from potential theory.  First, a circle of radius approximately twice the typical radius of $\Omega$ is placed 
concentric with it.
From potential theory, we know that any field generated by sources
outside of this circle can be approximated arbitrarily well by placing enough
equivalent charges on the circle.
In practice, it is sufficient to place a small number of ``proxy'' points spaced evenly on the circle.
For the relatively low frequencies (i.e. small $\omega$) tested in this paper,
we have found it is enough to have $75$ proxy points.
We call all the points on the neighboring obstacles that are within the proxy circle \textit{near} points.

\begin{figure}[ht]
\centering
\setlength{\unitlength}{1mm}
\begin{picture}(100,50)
\put(-7,00){\includegraphics[height=60mm]{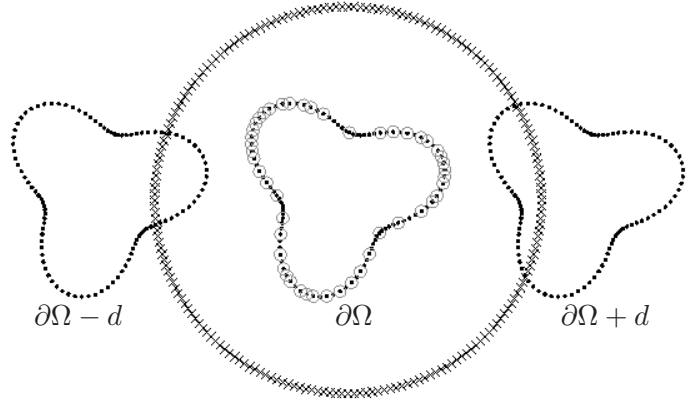}}
\put(50,13){$\partial \Omega$}
\put(10,13){$\partial \Omega-d$}
\put(80,13){$\partial \Omega+d$}
\end{picture}
\caption{\label{fig:sample} Illustration of proxy points ($\times$) and the skeleton points ($\circ$).  The near points are the points inside the proxy circle not belonging to $\Omega$.}
\end{figure}

Instead of computing the interpolatory decomposition of each $\mtx{A}_j$, one matrix $\mtx{P}$ is generated 
by computing the interpolatory decomposition of the matrix 
$$\left[\mtx{A}_j(:,I^{\rm near})| A^{\rm proxy})\right]$$
where $I^{\rm near}$ corresponds to the indices of the points on $\pO+j\dd$ that are near $\pO$, and $A^{\rm proxy}$ is a matrix that 
characterizes the interaction between the nodes on $\partial \Omega$ and the proxy points.  We call the points on $\pO$ picked by 
the interpolatory decomposition \textit{skeleton points}.  Figure \ref{fig:sample} illustrates the proxy points, near points and skeleton points for a 
sample domain.  This $\mtx{P}$ can be used for all $2P$ nearest neighbors.  For each $j$, the $\mtx{R}_j$ is simply given by $\mtx{R}_j = \mtx{A}_j(J(1:l),:)$.  

\begin{remark}
When $\mtx{A}$ is replaced by $\tilde{\mtx{A}}$
in \eqref{eq:blocksolve}, and the Woodbury formula used for $\tilde{\mtx{A}}$,
the number of matrix-vector
multiplies with $\mtx{A}^{-1}$ required to evaluate
\eqref{eq:blocksolve} is then $2M + q + 2Pl$.
\label{r:multiangle2}
\end{remark}

\section{Creating a compressed inverse of the matrix $\mtx{A}$}
\label{sec:HSS}
While the matrix $\mtx{A}$ is dense,
it has a structure that we call \textit{Hierarchically Block Separable (HBS)} which 
allows for an approximation of its inverse to computed rapidly.
Loosely speaking, its off-diagonal blocks are low rank.
This arises because $\mtx{A}$ is the discretization on a curve of an
integral operator with smooth kernel
(when $\omega$ is not too large).
This section briefly describes the HBS property and how it can be exploited to rapidly construct an approximate inverse 
of a matrix. For additional details see \cite{m2011_1D_survey}.
Note that the HBS property is very similar to the concept of \textit{Hierarchically
Semi-Separable (HSS)} matrices \cite{2007_shiv_sheng,gu_divide}.

\subsection{Block separable}
Let $\mtx{M}$ be an $mp\times mp$ matrix that is blocked into $p\times p$ blocks,
each of size $m\times m$.

We say that $\mtx{M}$ is ``block separable'' with ``block-rank'' $k$
if for $\tau = 1,\,2,\,\dots,\,p$, there exist $n\times k$
matrices $\mtx{U}_{\tau}$ and $\mtx{V}_{\tau}$ such that each off-diagonal
block $\mtx{M}_{\sigma,\tau}$ of $\mtx{M}$ admits the factorization
\begin{equation}
\label{eq:yy1}
\begin{array}{cccccccc}
\mtx{M}_{\sigma,\tau}  & = & \mtx{U}_{\sigma}   & \tilde{\mtx{M}}_{\sigma,\tau}  & \mtx{V}_{\tau}^{*}, &
\quad \sigma,\tau \in \{1,\,2,\,\dots,\,p\},\quad \sigma \neq \tau.\\
m\times m &   & m\times k & k \times k & k\times m
\end{array}
\end{equation}

Observe that the columns of $\mtx{U}_{\sigma}$ must form a basis for
the columns of all off-diagonal blocks in row $\sigma$, and
analogously, the columns of $\mtx{V}_{\tau}$ must form a basis for the
rows in all the off-diagonal blocks in column $\tau$. When (\ref{eq:yy1})
holds, the matrix $\mtx{M}$ admits a block factorization

\begin{equation}
\label{eq:yy2}
\begin{array}{cccccccccc}
 \mtx{M} &  =& \mtx{U}&\tilde{\mtx{M}}& \mtx{V}^{*} & +&  \mtx{D},\\
mp\times mp &   & mp\times kp & kp \times kp & kp\times mp && mp \times mp\\
\end{array}
\end{equation}

where
$$
\mtx{U} = \mbox{diag}(\mtx{U}_{1},\,\mtx{U}_{2},\,\dots,\,\mtx{U}_{p}),\quad
\mtx{V} = \mbox{diag}(\mtx{V}_{1},\,\mtx{V}_{2},\,\dots,\,\mtx{V}_{p}),\quad
\mtx{D} = \mbox{diag}(\mtx{D}_{1},\,\mtx{D}_{2},\,\dots,\,\mtx{D}_{p}),
$$

and

$$\tilde{\mtx{M}} = \left[\begin{array}{cccc}
0 & \tilde{\mtx{M}}_{12} & \tilde{\mtx{M}}_{13} & \cdots \\
\tilde{\mtx{M}}_{21} & 0 & \tilde{\mtx{M}}_{23} & \cdots \\
\tilde{\mtx{M}}_{31} & \tilde{\mtx{M}}_{32} & 0 & \cdots \\
\vdots & \vdots & \vdots
\end{array}\right].
$$

Once the matrix $\mtx{M}$ has put into block separable form, its inverse is given by 
\begin{equation}
\label{eq:woodbury}
\mtx{M}^{-1} = \mtx{E}\,(\tilde{\mtx{M}} + \hat{\mtx{D}})^{-1}\,\mtx{F}^{*} + \mtx{G},
\end{equation}
where
\begin{align}
\label{eq:def_muhD}
\hat{\mtx{D}} =&\ \bigl(\mtx{V}^{*}\,\mtx{D}^{-1}\,\mtx{U}\bigr)^{-1},\\
\label{eq:def_muE}
\mtx{E}  =&\ \mtx{D}^{-1}\,\mtx{U}\,\hat{\mtx{D}},\\
\label{eq:def_muF}
\mtx{F}  =&\ (\hat{\mtx{D}}\,\mtx{V}^{*}\,\mtx{D}^{-1})^{*},\\
\label{eq:def_muG}
\mtx{G}  =&\ \mtx{D}^{-1} - \mtx{D}^{-1}\,\mtx{U}\,\hat{\mtx{D}}\,\mtx{V}^{*}\,\mtx{D}^{-1}.
\end{align}

\subsection{Hierarchically Block-Separable}
Informally speaking, a matrix $\mtx{M}$ is \textit{Hierarchically Block-Separable} (HBS),
if it is amenable to a \textit{telescoping} version of the above
block factorization.  In other words,
in addition to the matrix $\mtx{M}$ being block separable, so is $\tilde{\mtx{M}}$
once it has been reblocked to form a matrix with $p/2 \times p/2$ blocks,
and one is able to repeat the process in this fashion multiple times.


For example, a ``3 level'' factorization of $\mtx{M}$ is
\begin{equation}
\label{eq:united4}
\mtx{M} = \mtx{U}^{(3)}\bigl(\mtx{U}^{(2)}\bigl(\mtx{U}^{(1)}\,\tilde{\mtx{M}}^{(0)}\,(\mtx{V}^{(1)})^{*} + \mtx{B}^{(1)}\bigr)
(\mtx{V}^{(2)})^{*} + \mtx{B}^{(2)}\bigr)(\mtx{V}^{(3)})^{*} + \mtx{D}^{(3)},
\end{equation}
where the superscript denotes the level.

The HBS representation of an $N\times N$ matrix requires $O(Nk)$ to store and to apply to a vector.
By recursively applying formula (\ref{eq:woodbury}) to the telescoping factorization, an approximation of the inverse can be computed with $O(Nk^2)$ computational cost; see \cite{m2011_1D_survey}.
This compressed inverse can be applied to a vector (or a matrix) very rapidly.
Note that for memory movement reasons, it is more efficient
to apply $\mtx{A}^{-1}$ to a block of vectors than
to each vector sequentially.

\section{Transmission problem}
\label{sec:trans}
So far this paper has focused on the Dirichlet problem.
However, all of the techniques presented here also apply to the
transmission problem, with minor modifications that we describe
briefly in this section.

\subsection{The boundary value problem}
This section presents the unit-cell version of the boundary value problem
analogous to section~\ref{s:restr}.

Let all obstacles in the grating have refractive index $n$,
and the background index be 1.
Then the partial differential equation analogous to \eqref{pde} is
\bea
(\Delta + n^2\om^2)u &=& 0 \qquad \mbox{ in } \Omega
\label{pdein}
\\
(\Delta + \om^2)u &=& 0 \qquad \mbox{ in } U\backslash\overline{\Omega}
\label{pdeout}
\eea
with matching conditions on the boundary analogous to \eqref{bc},
\bea
u^+ - u^- &=& -\ui \qquad \mbox{ on } \pO
\label{bcm}
\\
u_n^+ - u_n^- &=& -\uni  \qquad \mbox{ on } \pO
~.
\label{bcmn}
\eea
The quasi-periodicity and radiation conditions are the same as for the Dirichlet case.
Problems of this kind correspond to the acoustic scattering from obstacles with a
constant wave speed differing from the background value, or
electromagnetic scattering from a dielectric grating in TM
polarization. Remark ~\ref{r:guide} also applies.

\subsection{The integral formulation}
The transmission obstacle case is handled via an integral formulation
often called the M\"uller--Rokhlin scheme \cite{rokh83}:
a pair of unknown densities $\eta=[\sigma;\tau]$
(where $\sigma$ is single-layer and $\tau$ a double-layer)
is used on $\pO$,
and these densities also generate potentials {\em inside} $\Omega$ with
the interior wavenumber $n \omega$.
The resulting operator
$A$ turns out to be a compact perturbation of the 2-by-2 identity.
Operators
$\hat B$ and $\hat C$ become slightly more elaborate, whilst $\hat Q$
is unchanged. The complete recipe is found in \cite{qpsc}.

\subsection{The linear system}
As with the Dirichlet problem, $\partial \Omega$ is discretized via a Nystr\"om method with $N$ points, and the 
Sommerfeld contour is discretized as before with $M$ points.  However, since there is a pair 
of densities $\sigma$ and $\tau$ defined on the boundary (one of each for every point), the resulting 
linear system has $2N+2M$ unknowns.
In order for the matrix $\mtx{A}$ to be amenable to the fast direct solver described in section \ref{sec:HSS},
it is crucial to interlace the unknowns so that nearby entries in the vector have nearby locations on $\pO$,
namely $\vct{\eta} = [\sigma_1;\tau_1;\sigma_2;\tau_2; \cdots ; \sigma_N;\tau_N]$.
This reorders the columns of $\mtx{A}$; the same reordering is also needed for the rows.


\section{Numerical examples}
\label{sec:numerics}
In this section, we illustrate the performance of the proposed solution methodology for several different problems.
Once the obstacle, period $d$, frequency $\omega$, and
set of desired incident angles
are specified, the solver requires two steps: 

\bi
\item {\em Pre-computation:} A compressed representation of the matrix $\mtx{A}^{-1}$ 
is computed via the technique in section \ref{sec:HSS}.  If $P>0$,
the low-rank neighbor
contributions $\mtx{L}_j$ and $\mtx{R}_j$
are computed as in section \ref{sec:neigh}.
This step is performed once.

\item {\em Block solve:}
For each distinct Bloch phase $\alpha$ derived from the set of incident angles,
a $N$-by-$q$ right-hand side matrix is formed by stacking
the right-hand sides $\vct{b}$ in \eqref{eq:block}
for the $q$ incident angles that share this common $\alpha$.
The block solve formula \eqref{eq:blocksolve}
from Section \ref{sec:block} is then applied to this right-hand side matrix,
at a cost of $(q + 2M + 2Pl)$ matrix-vector products
with the compressed inverse.
\ei

It is advantageous for the user to choose many incident
angles that share a common $\alpha$.  This leads
an overall efficiency gain of a factor $q$ results (assuming $q \ll 2M+2Pl$).

All experiments are run on a Lenovo laptop computer with 8GB of RAM and a 2.6GHz Intel i5-2540M processor. 
The direct solver was run at a requested relative precision of $10^{-10}$.
It
was implemented rather crudely in MATLAB, which means that significant further
gains in speed should be achievable.

\begin{remark}[error measure]
In the below, the solution error quoted is the ``flux error'',
namely the absolute difference between the total incoming and outgoing flux.
Since these fluxes should be equal, this provides a standard
error measure in diffraction problems.
It is computed via a weighted sum of the squared magnitudes of the
Bragg amplitudes $c_n$ and $d_n$ in \eqref{rbu}--\eqref{rbd};
see \cite[Eq.~(4.2)]{qpsc}.
We have checked that this measure is similar to the pointwise error in the
far field.
\end{remark}

\subsection{Scaling with problem size}
\label{sec:scaling}
\begin{figure}[ht]
\centering
\setlength{\unitlength}{1mm}
\begin{picture}(100,65)
\put(-30,-5){\includegraphics[width=85mm]{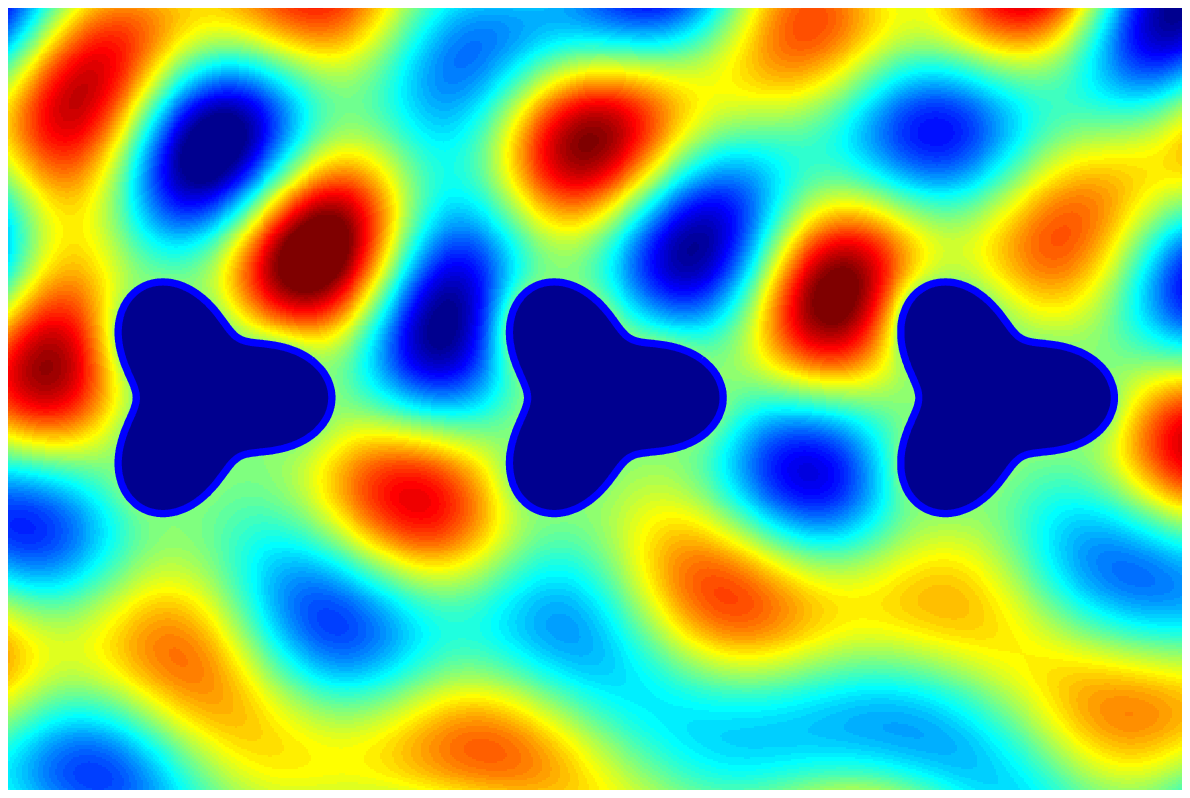}}
\put(50,-5){\includegraphics[width=85mm]{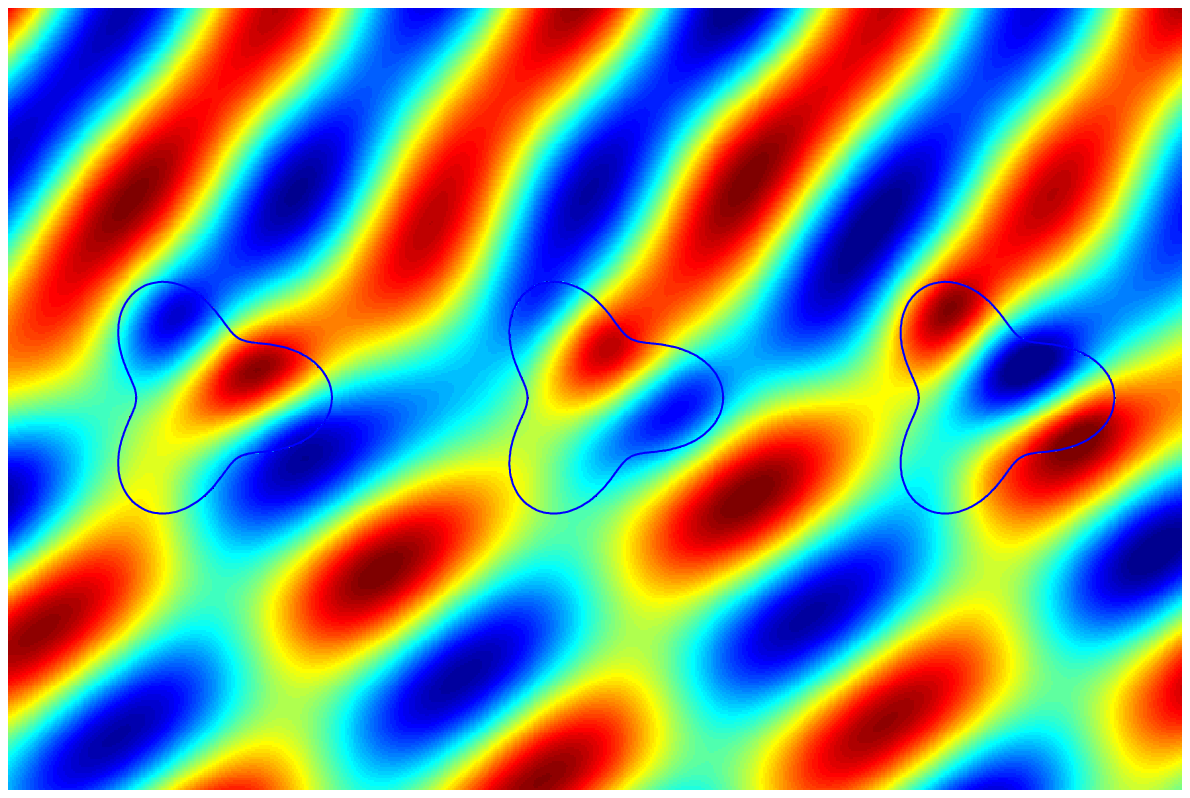}}
\put(10,00){(a)}
\put(90,00){(b)}
\end{picture}
\caption{\label{fig:stars} Illustration of the total field for star shaped obstacles with (a) Dirichlet and (b) transmission 
boundary conditions.}
\end{figure}

In this section, we apply the direct solver to both the Dirichlet and transmission scattering problems from a grating of star-shaped domains with period $d = 1$, and the frequency $\omega$ is fixed at $10$.
For the transmission case, the index is $n=1.5$.
We choose a single incident angle $\theta = -\pi/5$.  Hence, $q=1$.  The obstacle is 
defined by the parametrization $\zz(t) = (f(t) \cos t, f(t) \sin t)$
where the radius function is $f(t)= 0.35+0.105\cos(3t)$ for
angle $t\in [0,2\pi)$.
Figure \ref{fig:stars} plots the total field for both cases.

\begin{figure}[ht]
\centering
\setlength{\unitlength}{1mm}
\begin{picture}(100,70)
\put(-20,01){\includegraphics[width=75mm]{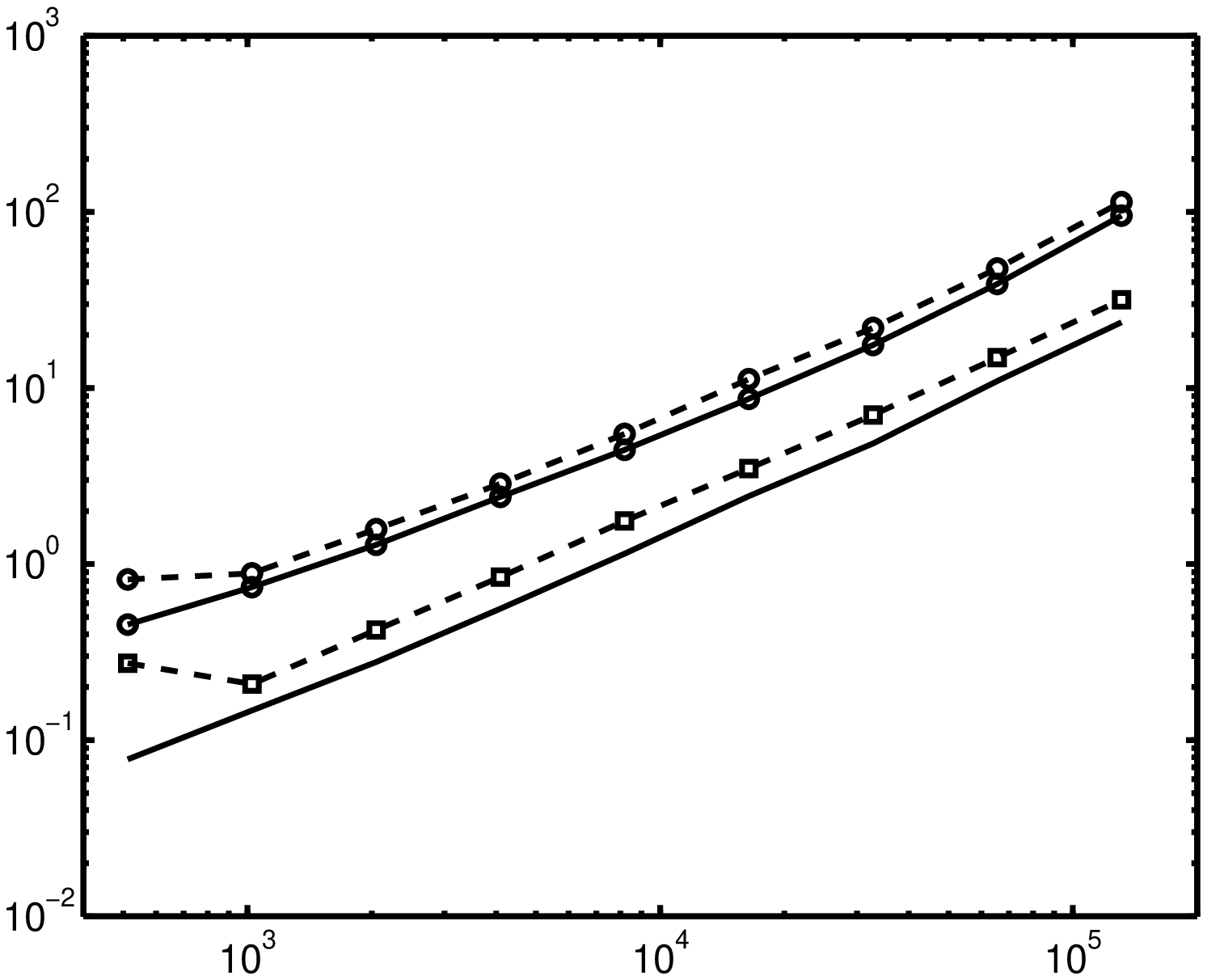}}
\put(50,01){\includegraphics[width=75mm]{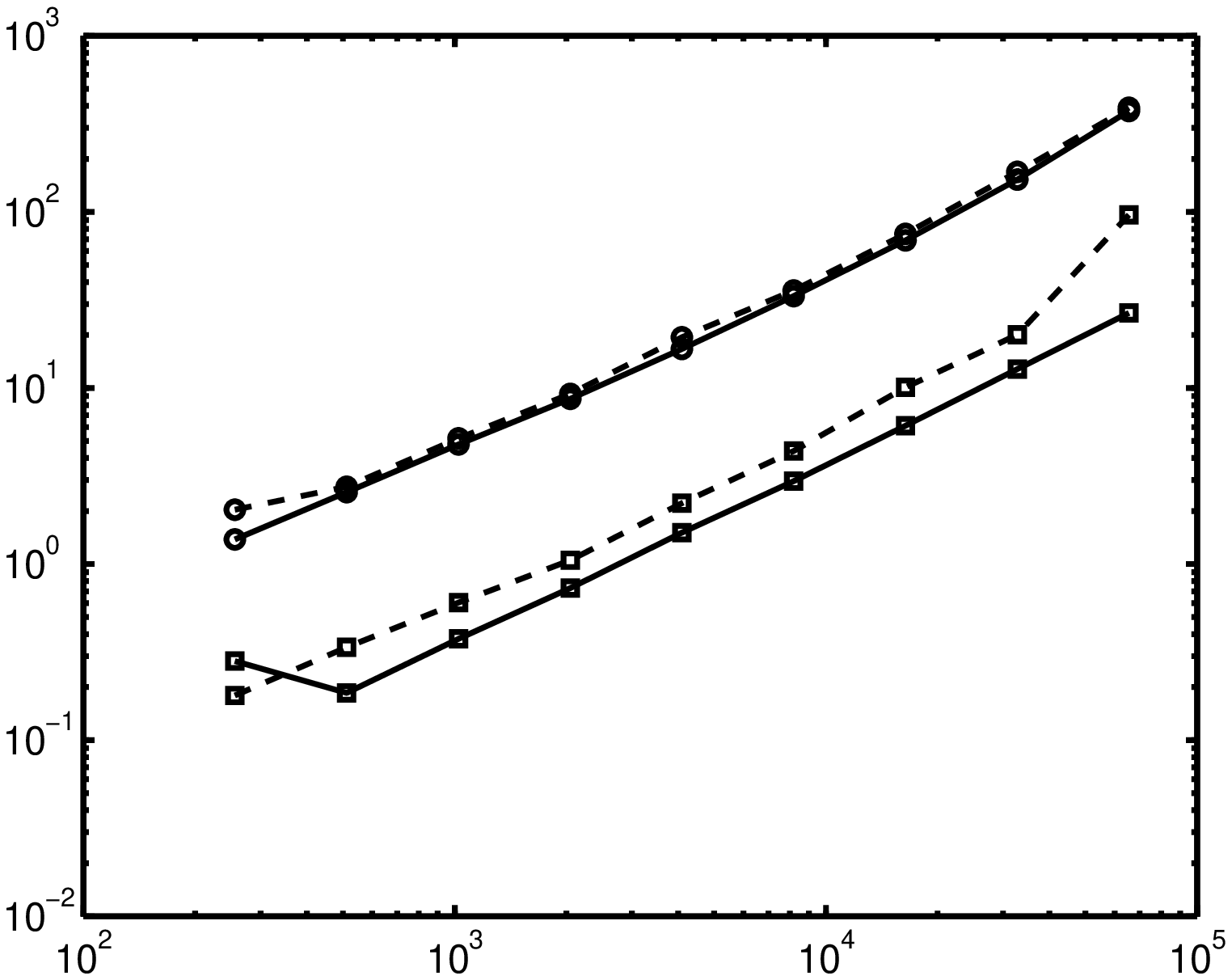}}
\put(-18,15){\rotatebox{90}{Time in seconds}}
\put(15,-3){(a)}
\put(15,01){$N$}
\put(85,-3){(b)}
\end{picture}
 \caption{\label{fig:times} Time in seconds vs $N$ for the (a) Dirichlet and (b) transmission boundary value problems.  
 The times for pre-computation are marked with $\circ$ while times for the block solves are reported with $\square$.  
The dash lines correspond to the times with contributions from neighbors while the solid corresponds to the times 
when no neighbor contribution is added. }
\end{figure}

The number of periodizing unknowns were kept fixed, with $M = 90$,
while the number of discretization points on the boundary 
of the obstacle was increased.  This experiment is designed to illustrate the scaling of the fast direct solver.  
(Note that for most of the $N$ tested, the boundary is over-discretized:
for $N = 512$, the discretized integral equation already has accuracy $10^{-8}$.)
Figure \ref{fig:times} illustrates the times for the pre-computation and the solve steps in the direct solver 
for the Dirichlet and transmission problems both with and without neighbor contributions ($P=1$).
 As expected, since $\pO$ is not space-filling and $\omega$ is small, the times scale linearly with $N$.  Notice
that the cost of adding the contributions from nearest neighbors is small, since their interaction rank is $l = 58$.
For the Dirichlet problem with $N = 2^{17}=131072$, 
it takes $95$ seconds for the precomputation and $24$ seconds for the block solve when $P=0$, versus 
 $114$ seconds for the precomputation and $32$ seconds for the block solve when $P = 1$.

\subsection{Comparison of fast direct solver against a fast iterative solver}
\label{sec:compare}
\begin{figure}{\includegraphics[height=90mm]{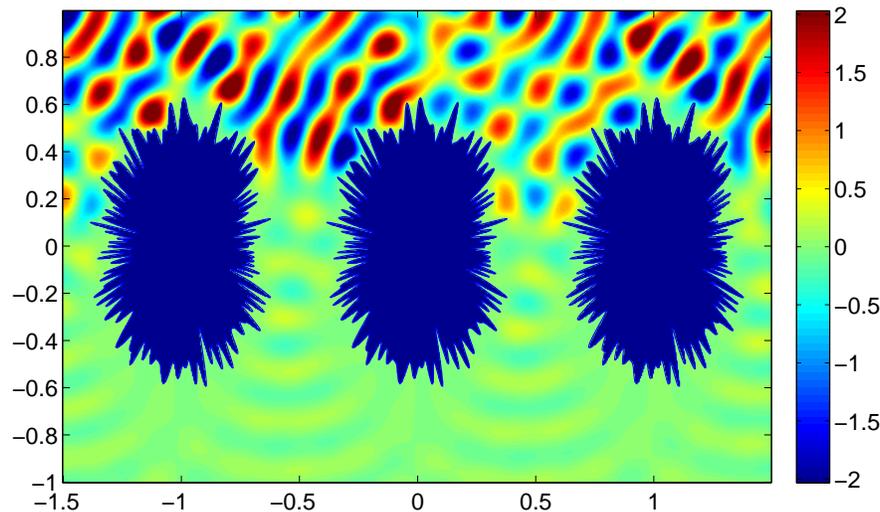}}\caption{\label{fig:woods}Illustration of the total field at a Wood's anomaly off a collection of complicated obstacles.} \end{figure}

In this experiment, we consider a more challenging Dirichlet problem with $\omega = 30$ where the obstacle $\Omega$ is 
the complicated domain whose total field is illustrated in Figure \ref{fig:woods}.
This domain is given by a radial function $f(t)$
which is a random Fourier series with 201 terms.
The first cosine term is set large and negative to give the obstacles
the shape of a vertically-oriented ellipse.
Its rough surface resembles a dentritic metallic particle.
The large number of tight curves and close approaches of the
boundary with itself
is typical of complex geometries, and demands a large $N$ to reach
any reasonable accuracy.
It also causes ill-conditioning that demands a large number of
GMRES iterations in an iterative solver.

The grating period is $d=1$, corresponding to around 5 wavelengths.
The closest distance between obstacles is around $0.25$.
We choose an incident angle corresponding to a Wood's anomaly ($\theta = -\arccos(1 - 2\pi/\om)$);
note that therefore the problem cannot even be solved the standard
integral-equation approach based on the quasi-periodic Green's function.
We take $P = 1$.  
In order to obtain an accuracy of $10^{-8}$ for this problem, $10^5 $ unknowns are needed on the boundary 
of the obstacle ($N$) and the number of periodizing unknowns was increased to $M = 120$.

Two methods were used to solve for the unknown densities.
Firstly we tested an iterative solution of \eqref{eq:block2}
using standard GMRES without restarts, with the publicly-available Helmholtz
FMM of Greengard--Gimbutas \cite{HFMM2D} to apply the matrices
in the $\tilde{\mtx{A}}$ block (with quadrature corrections near the
diagonal where appropriate), and dense matrix-vector multiplication
for the other three blocks.
Secondly we tested the direct solver scheme presented in this work.
Note that the FMM is coded in fortran,
whereas our direct solver is (apart from the interpolative decomposition)
in MATLAB.

The GMRES+FMM solver takes
approximately one hour, taking a large number ($248$) of iterations,
to solve for the densities. By using the fast direct solver, the densities 
can be found with $4.5$ minutes for the pre-computation and $50$ seconds for the block solve at each $\alpha$.  In other
words, for this domain, the direct solver can solve $66$ independent incident angles in the
amount of time it takes the accelerated iterative method to solve for one.  

Note that for the neighbor interactions, we chose 200 points on a proxy circle with diameter $1.1$ times
the vertical height of the obstacle.
The approximate interaction rank between 
obstacles is then $l = 139$.  Note a single matrix vector multiply
with $\mtx{A}^{-1}$ takes $3$ seconds.  Thus if we were to apply $\mtx{A}^{-1}$ to 
one vector at a time in \eqref{eq:inv} it would take $3(2M + 2Pl) = 1554$ seconds.  
Thus, since the complete block solve takes $50$ seconds, significant timing gains 
are seen by applying the $\mtx{A}^{-1}$ to block matrices instead of single vectors.

Next, we present an example where
a large sampling of incident angles are required,
as would be typical for a solar-cell design problem.
An arithmetic series in
$\cos\theta$, with around 200 values 
covering the range $(-1,1)$, is considered.
Their spacing in $\cos\theta$ is $2\pi/(21\, \om d)$,
which provides roughly $21\,q$ incident angles,
where $q = \om d/\pi \approx 10$.
As discussed above, because around $q$ angles share the same $\alpha$,
this leads to an additional speed-up of
a factor of nearly $q$.
It takes $19.1$ minutes to solve for the $200$ densities
($4.1$ minutes of pre-computation, followed by $15$ minutes for the block solves.)
Notice that this is around 600 times faster than the solution at 200 incident
angles would take using GMRES+FMM.

\begin{figure}[ht]
\centering
\setlength{\unitlength}{1mm}
\begin{picture}(120,65)
\put(-12,03){\includegraphics[width=80mm]{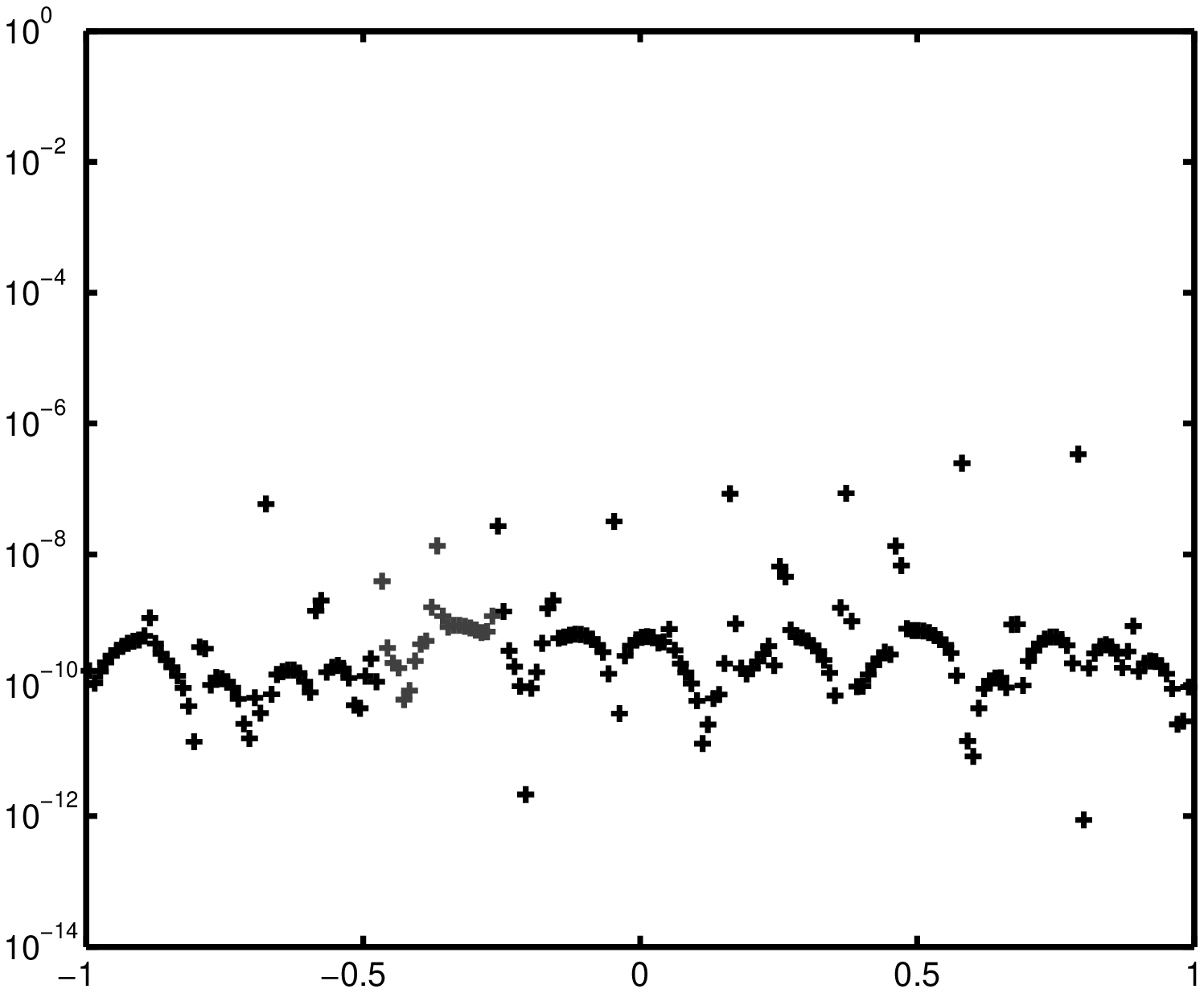}}
\put(70,03){\includegraphics[width=80mm]{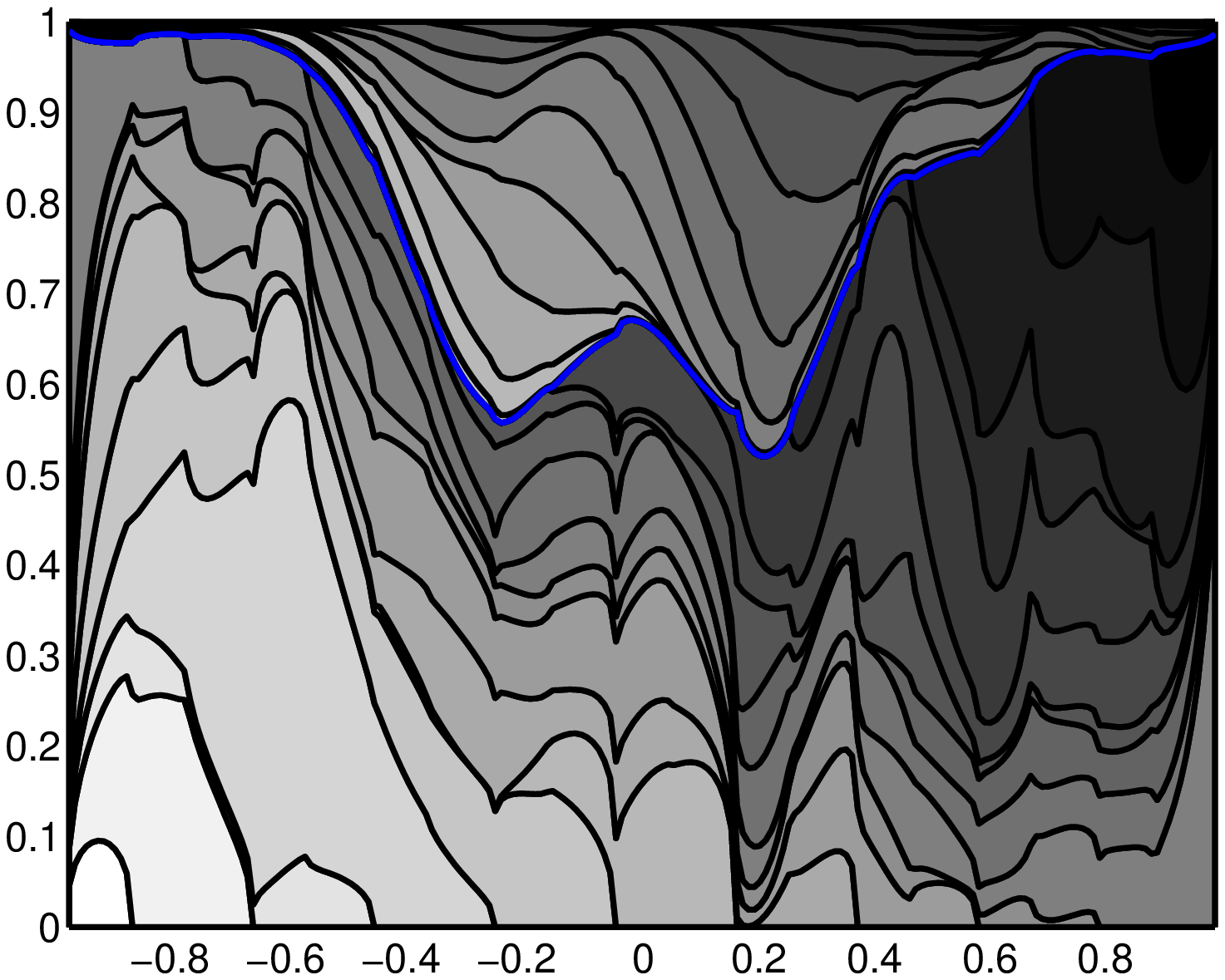}}
\put(27,00){(a)}
\put(25,05){$\cos \theta$}
\put(-10,25){\rotatebox{90}{Flux error}}
\put(70,15){\rotatebox{90}{Cumulative flux fraction}}
\put(110,00){(b)}
\end{picture}
\caption{\label{fig:multi_ang}   (a) The flux errors and (b) the outgoing flux fractions for all Bragg modes when solving a 
Dirichlet problem with $200$ incident angles and $\om = 30$.
In (b) each flux fraction is shown in a different gray shade,
and the solid blue line separates reflected from transmitted intensity.}
\end{figure}

The resulting fractions of incident flux scattered into each of the Bragg modes
(i.e.\ $|c_n|^2$ and $|d_n|^2$) are shown, as a function of incident
angle $\theta$, in Fig.~\ref{fig:multi_ang} (b).

\subsection{A challenging transmission problem}

\begin{figure}{\includegraphics[height=100mm]{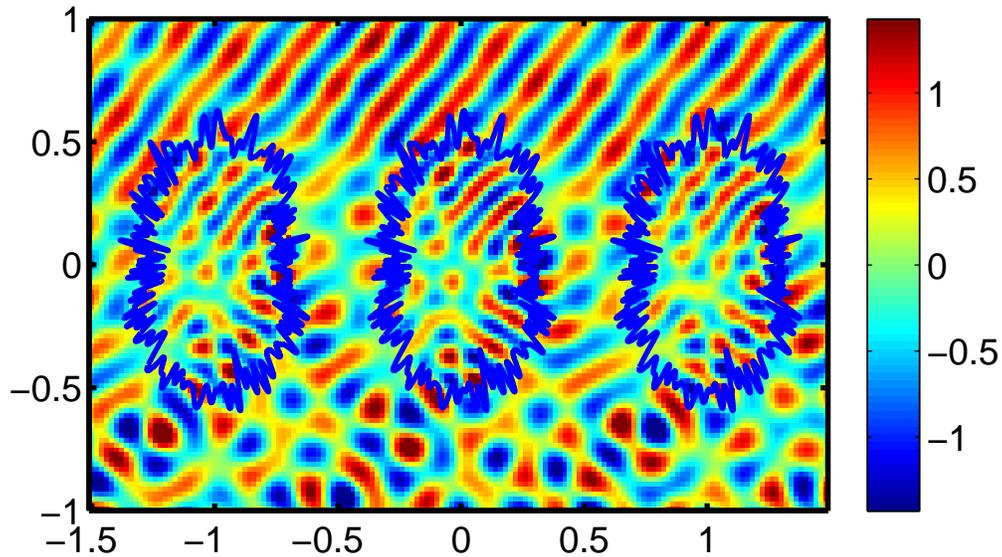}}\caption{\label{fig:transfield}Illustration of the total field at a Wood's anomaly off a collection of complicated obstacles with 
transmission boundary conditions.} \end{figure}

Prior to the development of the fast direct solver, accurately solving a transmission problem on complicated domains required 
a lot of computing time.  We applied the direct solver to a transmission problem on the domain from section \ref{sec:compare}.  
With $N = 10^4$
and all other parameters as in the previous example
(thus there are 20241 unknowns),
the solver achieves a flux error of $10^{-5}$ in $6$ minutes of 
pre-computation and $14$ seconds for each block solve. Figure \ref{fig:transfield} illustrates the total field for this example.

\section{Conclusion}
\label{sec:conc}
This paper presented a fast direct solution technique for grating scattering problems with either 
Dirichlet or transmission boundary conditions.  For low frequency problems on simple domains, 
the computational cost of the solver scales linearly with the number of discretization points 
on one obstacle.  The example in section \ref{sec:compare} illustrates that 
when the obstacle is complicated and the frequency somewhat higher,
the direct 
solver is much faster than an FMM accelerated GMRES, because
it handles ill-conditioning.
Additionally, 
the direct solver is very fast for multiple incident angles that occur often in design problems,
and this can be further accelerated in the case when many incident
angles share a Bloch phase $\alpha$.
In one complicated Dirichlet obstacle case which
requires $10^5$ unknowns to discretize, 200 incident angles
are solved in under 6 seconds per incident angle.

Although, for simplicity, we disallowed intersections of
$\pO$ with the $L$ and $R$ walls,
it would be quite simple to adapt the fast direct solver to
the scheme presented in \cite[Sec.~6]{qpsc} to handle this case.
It would also be relatively easy to generalize the scheme to multi-layer
transmission gratings, which are more common in applications.
We anticipate creating a fast solver for this case in future work.


\bibliographystyle{abbrv} 
\bibliography{alex}

\end{document}